# Efficient Coherent XUV Generation and Manipulation in Microfluidic Glass Devices

A. G. Ciriolo[1,*], R. Martínez Vázquez[1,*], G. Crippa[1,2], M. Devetta[1], D. Faccialà[1], P. Barbato[1,2], F. Frassetto[3], M. Negro[1], F. Bariselli[5],

L. Poletto[3], V. Tosa[4], A. Frezzotti[5], C. Vozzi[1, §], R. Osellame[1], S. Stagira[1, 2]

[1] *National Research Council (CNR), Institute for Photonics and Nanotechnologies, Milano, Italy*
[2] *Politecnico di Milano, Physics Department, Milano, Italy*
[3] *National Research Council (CNR), Institute for Photonics and Nanotechnologies, Padova, Italy*
[4] *National Institute for R&D of Isotopic and Molecular Technologies, Cluj-Napoca, Romania*
[5] *Politecnico di Milano, Department of Aerospace Science and Technology, Milano, Italy*

*\*these authors contributed equally to this work*

[§] e-mail: caterina.vozzi@ifn.cnr.it

*Abstract*—The development of compact and bright XUV and soft X-ray sources based on high-order harmonic generation is boosting advances towards understanding the behavior of matter with extreme temporal and spatial resolutions. Here, we report efficient XUV generation inside microfluidic devices fabricated by femtosecond laser irradiation followed by chemical etching. Our microfluidic approach allows one to control and manipulate the generation conditions in gas on a micro-meter scale with unprecedented flexibility, thus enabling a high photon-flux and broadband harmonics spectra up to 200 eV.

## I. Introduction

Nowadays, table-top ultrafast sources of coherent eXtreme UltraViolet (XUV, 10-124 eV) and soft-X ray (from 124 eV up to few keV) radiation based on High-order Harmonic Generation (HHG) are noticeably contributing to scientific advances in understanding the behaviour of matter upon electronic excitation on an ultrashort time scale [1-7]. The key point of the success of these sources relies on the unique capability to combine extreme temporal and spatial resolution, allowing the study of ultrafast dynamics with atomic specificity and chemical environment sensitivity down to the attosecond temporal domain (1 as = $10^{-18}$ s).

Besides the astonishing potential for unveiling matter dynamics on an extreme time scale, the HHG technology is still undergoing continuous progress, intended to overcome several fundamental limits that significantly hinder its applications.

Namely, the dramatically low conversion efficiency of HHG still represents a major issue, especially in the soft-X-ray domain, wherein mid-infrared (mid-IR) driving pulses must be used to drive the conversion process. Indeed, HHG shows an extremely unfavourable dependence of the yield on the driving pulse wavelength, which scales as $\lambda^{-n}$ with $5 \leq n \leq 7$.

In this framework, remarkable attention has recently been paid to the development of intense ultrafast near/mid-IR driving sources at high repetition rates for increasing the photon fluence [8,9] e. On the other hand, a great effort is currently being devoted to boosting the performances of the HHG sources towards brighter XUV and soft-X ray emission. Several different configurations have been proposed for enhancing the conversion yield *per* driving pulse, including HHG in gas-filled capillaries [10] and gas cells [11].

Capillaries are routinely used in the field of non-linear optics and ultrafast sources for confining the laser beams over longer distances than the Rayleigh range. For instance, gas-filled hollow-core fibers are widely exploited in nonlinear optics [12] and for the compression of intense laser pulses [13]. By combining the benefit of a long generation medium and a high-energy parametric source, an extremely bright supercontinuum generation from the XUV up to the keV photon energy was demonstrated by phase-matched emission in a gas-filled capillary driven by millijoule-level mid-IR ultrashort pulses [14].

For HHG, capillary-based systems are commonly operated in a hydrostatic regime, such that uniform gas densities or shallow pressure gradients can be implemented. The hydrostatic regime ensures a high conversion yield by allowing emission from atoms all along the capillary [14-17]. However, this configuration provides a limited control on phase-mismatch and re-absorption by the gas. Re-absorption of XUV radiation from the gas throughout the waveguide becomes extremely critical, and optimal conversion in a long and uniformly dense generation medium is attained over a distance on the scale of the absorption length, provided a coherence longer than the characteristic absorption ($L_{coh}>5\ L_{abs}$) [18]. Moreover, uniformly filled hollow fibers may induce several undesired processes, including long-lived plasma, detrimental for operation at high driving repetition rates [19], and phase-distortion of the fundamental pulses due to nonlinear processes like self-phase modulation.

To date, *Quasi Phase Matching* (QPM) by a periodic corrugation of the fiber inner diameter for modulating the laser peak intensity [20,21] provides the experimental demonstration of enhanced harmonic generation yield by control of the capillary geometry on a sub-millimeter scale.

However, the coherent growth of the XUV radiation in capillaries is sensitive to any perturbation of the driving field, due to radiative losses during the propagation and dispersion in the neutral and ionized gas and mode-beating [22]. Thus, the phase-matching condition might significantly change along the propagation [23], requiring a local adjustment of the periodicity to recover in-phase dipole emission. HHG in aperiodic configurations for increasing QPM emission of selected harmonics has been theoretically proposed [24-26], but not experimentally demonstrated yet, mainly because of the technical difficulties associated with the control and optimization of the waveguide microstructure.

In this framework, the remarkable advances in micromachining techniques offer novel perspectives to coherent nonlinear conversion processes. Indeed, the availability of microfluidic systems allows tailoring of both the capillary shape and gas density modulation, thus enabling control of the generation medium, well beyond the hydrostatic regime.

We propose a novel approach to XUV generation and manipulation based on microfluidic glass devices, which is intended to combine the benefits of microfluidics with ultrafast nonlinear optics in hollow-core fibers.

We exploit the Femtosecond Laser Irradiation followed by Chemical Etching (FLICE) micromachining technique for building embedded hollow three-dimensional structures in the bulk of glass substrates [27-29]. We use these microstructured devices as microfluidic systems for fine control of the gas density in the laser-gas interaction volume. Herein, we measured efficient XUV generation and we demonstrate the possibility to manipulate and control the phase-matching conditions by tailoring the gas density with micrometer accuracy. This approach is suitable for more complex engineering with unprecedented flexibility and marks a promising route to high-brightness tabletop ultrafast XUV sources.

## II. CONCEPT: MICROFLUIDIC SOURCES FOR HHG

The archetypal structure of our devices is composed of a hollow-core waveguide and a microfluidic module for gas delivery and manipulation integrated into a monolithic finger-top glass chip engineered to work in a vacuum environment (see figure 1a,b).

Here we demonstrate efficient HHG in two different designs of microfluidic devices.

The first design [30] is a down-scaling version of the capillary scheme. The gas, injected by the microfluidic module, flows through a few-mm-long hollow waveguide under the pressure gradient downstream of the capillary extremities. In this device, the flow is nearly isothermal, with the density profile being proportional to the pressure, which is smoothly varying from the center to the outputs of the waveguide, thus giving a central region with almost a constant density profile [31].

The second design is equipped with a more sophisticated gas-delivery module, composed of an array of identical De Laval micronozzles, whose supersonic outflow improves spatial focusing of the gas sources at the inlets of the main channel on the micrometer length scale. These micronozzles are directly interfaced with the hollow waveguide, such that *local* gas density peaks can be realized along the beam path. The tailoring of the gas density modulation in this kind of device can be engineered by changing the shape, the number, and the position of the micro-nozzles.

The coherent build-up of harmonics through QPM based on structured gas density distribution was theoretically proposed by Auguste *et al.* in 2007 [32]. The experimental demonstration of this effect has been reported by J. Seres *et al.* [33]. In their work, the modulation of the gas density was obtained by adding two gas sources in a focused laser beam. However, in a tight-focusing geometry, scaling this scheme to a higher number of gas sources is challenging. Furthermore, based on the theoretical work by Hadas *et al.* [34], on HHG in a shallow focusing regime, the nonlinear optical conversion by QPM in a periodically modulated gas medium may be strongly inefficient. This result suggests investigating the possibility to overcome the theoretical limitations of a periodical geometry by the use of a non-periodic gas density modulation.

HHG in hollow-waveguide integrated within a microfluidic device offers an ideal platform for exploring this scheme. Hence, we exploited FLICE micromachined devices to demonstrate, for the first time, the generation of harmonics in periodic and aperiodic gas density modulations through an array of multi-jets inside a hollow waveguide.

## III. REALIZATION AND CHARACTERIZATION OF THE MICROFLUIDIC SOURCES

The microfluidic glass devices are fabricated on $1 \times 10 \times 8$ mm$^3$ fused silica substrates (FOCtek Photonics). The standard design of a device consists of a 100-µm-deep rectangular-like thin reservoir ($2 \times 4.7 \times 0.1$ mm$^3$) from where an arrangement of microchannels (or nozzles) departs, distributing the gas into the hollow-core waveguide, which is 8 mm long and has a diameter of 130 µm. The waveguide is embedded into the glass at a 300-µm depth. The devices are designed to work under vacuum.

In the first design, the waveguide is filled with gas by four evenly distributed injection micro-pipes with cylindrical shape, arranged at a distance of 1.2 mm from each other to cover the entire area of the reservoir and to allow for a uniform gas distribution along the waveguide. An abrupt decrease of density takes place in the proximity of the waveguide edges due to a remarkable acceleration of the gas towards the vacuum environment.

In the second design, we changed the profiles of the microchannels for the gas distribution from cylinders to De Laval micronozzles, and also their spatial location along the waveguide. These micronozzles are characterized by a convergent-divergent profile with diameters of 220 µm, 60 µm, and 90 µm in the input, throat, and output, respectively. The convergent zone has a length of 55 µm and the divergent one of 75 µm. We realized two different multi-jet microfluidic sources, the first composed of four evenly distributed gas jets (relative distance 1.2 mm), the second composed of three gas jets accommodated at different relative distances (1.9 mm and 1.4 mm). In front of the nozzle outlets, we arranged an exhaust rectangular opening to allow free gas expansion and to reduce gas stagnation in between two adjacent gas jets. To preserve the optical structure of the hollow waveguide, the width of the opening was set at 90 µm, smaller than the waveguide diameter. A picture of the device with micronozzles is shown in fig. 1(a). The devices are fabricated by the FLICE technique, which consists of irradiating the sample by high repetition rate femtosecond pulses. In the focal volume, a high amount of energy is deposited, inducing a permanent modification of the material which becomes locally more sensitive to the chemical attack by an etchant solution.

The trajectory followed by the laser focus inside the sample defines the irradiation pattern to be etched. Thus, by precisely moving the sample relative to the beam focus, arbitrarily shaped cavities can be obtained with micrometer-sized accuracy.

The fabrication of the glass microchips begins with the irradiation of the fused silica slab with a focused femtosecond laser beam (Satsuma HP, Amplitude), with a repetition rate of 1 MHz, 230 fs pulse duration, and pulse energy of 300 nJ. We used the second harmonic (515 nm) of the laser beam and focused it inside the glass slab by a 63X microscope objective (0.65 NA) endowed with a correction ring (LD-Plan Neofluar, Zeiss). The sample is moved with respect to the laser beam thanks to a three-axis stage (200 nm resolution ANT stages, Aerotech), following the desired trajectory. Afterward, the sample is immersed for 2-3 hours into a 20% aqueous solution of hydrofluoric acid (HF) in an ultrasonic bath at 35 °C, to remove the irradiated regions and reveal the desired empty channels.

The main characteristic of FLICE technique is the possibility of combining in the same device three-dimensional empty structures with arbitrary profiles and different dimensions (from micrometer to mm), such as the micro-nozzles and the hollow waveguide. To succeed in it, the irradiated pattern has been engineered to compensate for inhomogeneous exposition to acid during the etching step. This allows achieving a central channel with a constant radius and De Laval nozzles with desired dimensions.

The Navier-Stokes-Fourier (NSF) [35] equations for a viscous and compressible fluid have been used to study the steady gas flow through the device. A laminar flow regime has been assumed, due to the relatively low value of the reference Reynolds number. No-slip boundary conditions have been applied to walls because of the small value of the Knudsen number [36] in the whole flow domain. NSF equations have been solved numerically on the commercial Comsol Multiphysics™ CFD platform [37]. Details about the adopted grid structure are given in the supplementary materials.

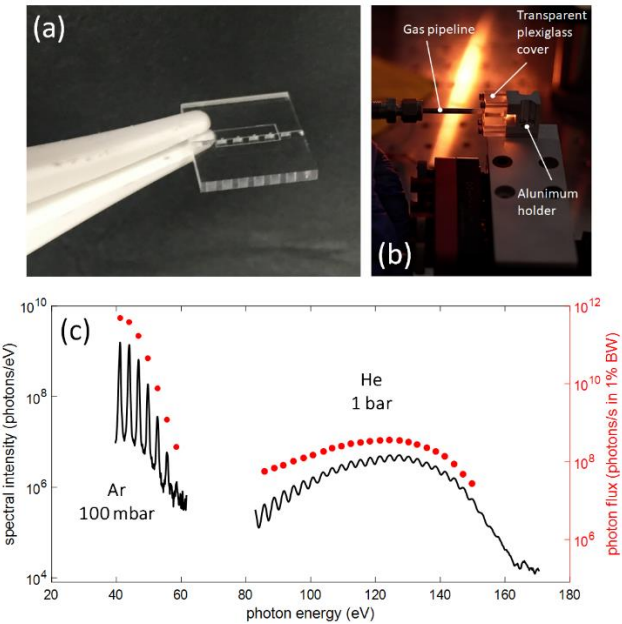

*Figure 1 XUV generation in microfluidic devices. (a) Picture of a microfluidic device for HHG. The mounting of the device is shown in figure (b). It is composed of aluminum holder and of plexiglass, the chip being encapsulated in-between. A viton gasket is used for sealing. The gas is delivered through a pipeline directly connected to the plexiglass cover. The mounting is placed on a highly-precision motorized transltional stage. (c) HHG spectra generated in a glass device composed of a 8-mm long capillary. The Argon spectrum is generated at a backing pressure of 100 mbar, the Helium spectrum at a backing pressure of 1 bar. On the right axis, in red, the corresponding photon flux per second in 1% bandwidth is reported. Picture (b) courtesy of Simone Campanella.*

IV. MICROFLUIDIC XUV SOURCE

To drive HHG, we used pulses at 800-nm wavelength delivered by an amplified Ti:Sapphire laser source (Amplitude, Aurora laser system: 15 mJ, 25 fs, 1 kHz). The laser pulses for this experiment were 30 fs long and with an energy of 500 µJ. The pulses are focused into the devices by a lens with a 30-cm focal length, matched for coupling the $EH_{11}$ hybrid mode [38]. The alignment of the device is performed by a high-precision multi-axes motorized translational system, inside a vacuum chamber. A picture of the HHG setup is shown in fig. 1(b).

The coupled mode at the output of the waveguide is monitored by an auxiliary beamline equipped with a mirror mounted on a motorized translational stage that can be inserted into the beamline and image the beam mode onto a beam profiler.

The gas is delivered to the device by a pipeline that is directly interfaced with the mechanical mounting that contains the chip. The gas density can be manually tuned by a needle valve mounted upstream the gas pipeline, and the backing pressure is accurately monitored by a capacitive pressure gauge placed after such valve. The working pressure in the generation chamber ranges between $10^{-4}$ and $10^{-5}$ mbar depending on the measured gas backing pressure. The grazing incidence XUV spectrometer used for collecting the HHG radiation works in stigmatic configuration [39]. The HHG signal is dispersed by a grazing incidence grating (Hitachi, working range 5-100 nm) and detected by a Micro-Channel Plate (MCP) followed by a phosphor screen. The image displayed on the phosphor screen is acquired by a CCD camera (Andor, Apogee Ascent A1050).

providing photon fluxes up to $6 \cdot 10^{12}$ photons/s in 1% bandwidth at 20-30 eV [47-49]. Besides HHG by single-pass on gas targets, multi-MHz cavity-enhanced HHG technology driven by high average power lasers has recently experienced

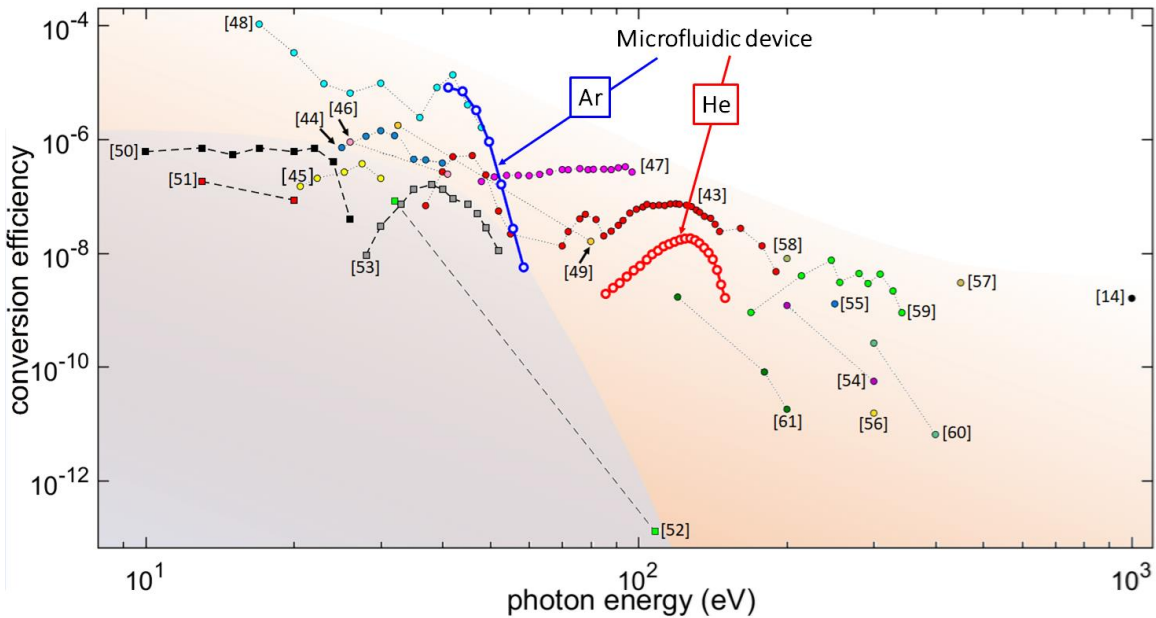

*Figure 2 State of the art in HHG-based XUV and soft-X ray sources. The graph shows the conversion efficiency (CE) in 1% bandwidth. If not directly provided, the 1%-BW CE was extimated from the data and experimental parameters reported in the manuscripts. Circoles mark single pass HHG, squares mark cavity enhanced HHG. The shaded areas defines the conversion efficiency range covered by single-pass (light-red) and cavity-based (light blue) sources.*

The photon flux was measured by a low-noise air-cooled CCD camera (Pixis-XO 2048B, Princeton Instrument) mounted in the focal plane of the spectrometer in place of the MCP, and working in the X-ray range between 10 eV and 30 keV. To remove the co-propagating fundamental beam, we used metallic filters (110-nm Al filter for HHG in Argon, 150-nm Pd filter for HHG in helium). The photon number per pulse is obtained by accounting for the absorption of the metal filters [40], the grating efficiency of the XUV spectrometer [41], and the quantum efficiency of the camera [42].

In figure 1(c), we report HHG spectra generated in Argon and Helium for the first design configuration. With Argon, the generation yield is maximized at a backing pressure of 100 mbar. In these conditions, harmonics up to 60 eV were generated. We measured peak photon flux approaching $10^{12}$ photons/s in 1% bandwidth at 41 eV, corresponding to a conversion efficiency of $8 \cdot 10^{-6}$ in 1% bandwidth. Compared to capillary-based XUV sources operated in similar experimental conditions (1-cm capillary filled with 120 mbar of Ar, 400 µJ Ti:Sa pulses at 1 kHz repetition rate) [43], the microfluidic source shows an enhanced photon flux by almost two orders of magnitude.

To date, higher photon fluxes, up to $10^{13}$ photons/s in 1% bandwidth, were generated in the spectral region 30-40 eV through HHG in continuous gas-jet driven by µJ-level high-repetition-rate sources, operating from tens of kHz to the 10 MHz [44-46]. Bright XUV generation in a gas cell driven by multi-millijoule 10-Hz driving laser sources was also reported, huge advances in photon fluxes delivery in the spectral range between 20 and 60 eV [50-53]. However, single-pass HHG still outperforms the efficiencies of cavity enhanced generation systems, mainly because of intracavity phase instabilities and detrimental cumulative plasma effects inside the generation volume that might hinder the in-cavity process [51].

To frame the performances of our microfluidic system with the state-of-the-art HHG XUV and soft-X ray sources, we compare the conversion efficiency (CE) *per* driving pulse energy as a function of the generated photon energies in fig. 2. By inspection, it results that the XUV generation efficiency in Argon is in line with that of the most efficient XUV sources reported in the literature so far.

Efficient HHG in Helium by the microfluidic source required higher backing pressure of 1 bar. The harmonics spectrum ranged from 80 eV up to 150 eV, with a photon flux larger than $10^7$ photons/s in 1% bandwidth within the whole spectral range, and a peak-value of $8 \cdot 10^8$ photons/s in 1% bandwidth at 130 eV, corresponding to a conversion efficiency of $1.8 \cdot 10^{-8}$ in 1% bandwidth. We observed an improvement of the photon flux by increasing the backing pressures, with a favorable indication of non-detrimental absorption. However, the pressure range we could access so far did not allow photon flux optimization in the microfluidic cell, since the maximum baking pressure was limited by the vacuum pumping system to 1 bar, at continuous gas flow. The device can operate in a multi-bar regime but a differential pumping is needed to apply pressures above 1 bar.

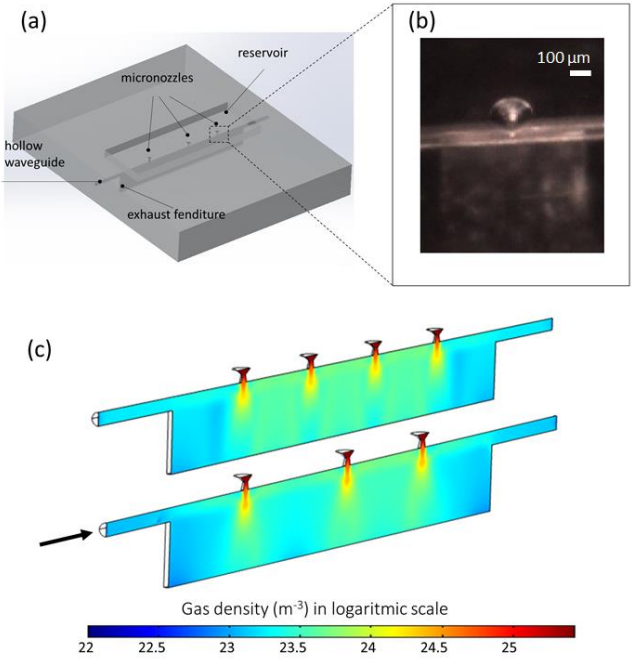

*Figure 4 HHG in the multi-jet devices. (a) Structure of a device with multi-jets for HHG. (b) Microscope view of a portion of the device where you can see the micronozzles and the hollow-core waveguide. (c) We report a sectional view of the simulated gas density ($m^{-3}$) in the devices with four and three nozzles. The black arrow indicates the laser entrance direction.*

For photon energies above 100 eV, near- and mid-IR (1.5 - 4 µm) ultrashort pulses are routinely exploited to drive HHG [14, 54-60]. This leads to a dramatic drop of the conversion efficiency of single pass schemes, due to the unfavourable scaling with fundamental driving wavelength. In this framework, scaling the system to higher gas densities is a fundamental requirement for efficient XUV emission in the water window [61]. Indeed, the suppression of the single-atom yield at long driving wavelength can be partially counterbalanced by a favorable phase-matching in the multi-bar regime and by an increase transparency at photon energies approaching the keV [10, 14], allowing enhanced conversion efficiency by coherent superposition from highly dense and long gas media.

In this sense, the microfluidic sources provide a suitable technology for enabling both beam confinement in hollow waveguides and high local gas densities. Furthermore, they ensure the unique capability to accurately control and manipulate the gas distribution for tailored phase-matching conditions, as shown in the next section.

## V. HHG WITH INTEGRATED ARRAY OF GAS JETS

In the second design, we aimed at obtaining sufficiently focused and confined gas jets, providing desidered gas density peaks along the wave guide axis. In both configurations (with three and four nozzles), we arranged the gas jets at a distance much larger than their transversal size in order to obtain a high gas density contrast between jets and background. The geometry was preliminarily optimized by running CFD simulations based on the fluid models and computational tools mentioned in Section III and produces a moderately supersonic gas flow inside the waveguide, localized at the output of the nozzles.

Figure 3(c) reports a sectional view of the numerical gas density distributions computed at a backing pressure in the reservoir of 1 bar. The gas distribution is composed of highly dense and confined gas jets interposed to the low-density background, with a high peak-to-background contrast ratio: the density contrast is >7 in the middle of the waveguide while it is >20 on the two extremities of the waveguide. Under these conditions, we can minimize the contribution to HHG from the gas background and we can study the dependence of the harmonic emission on the gas density modulation by directly comparing the harmonics spectra in different gas-jets configurations.

Figure 4 shows the experimental high-order harmonic spectra generated in the two devices. A clear difference between the two spectra can be noticed throughout the whole spectral range. Indeed, the yield is not monotonically related to the number of sources, but it exhibits a counterintuitive re-shaping. In the three-jet aperiodic configuration, we observe a cutoff extension associated with a local increase in the spectral yield at higher-harmonic orders. With the periodic four nozzles arrangement, the photon energy cutoff is located at 140 eV and is associated with a higher relative yield. With the aperiodic three-jets arrangement, we observed a remarkable extension of the harmonic spectrum up to 200 eV, with a cutoff at 190 eV.

To investigate the growth of the harmonic components at higher energies in a tailored gas density, we calculate the phase-mismatch $\Delta k_q$ between the fundamental and harmonic field for each harmonic order $q$. We used a one-dimensional approximation, in which we assumed the fundamental coupled to the $EH_{11}$ mode of the hollow waveguide. The pulse is modelled in the temporal domain as a Gaussian envelope with duration and intensity of 30 fs and $9.5 \cdot 10^{14}$ W/cm$^2$,

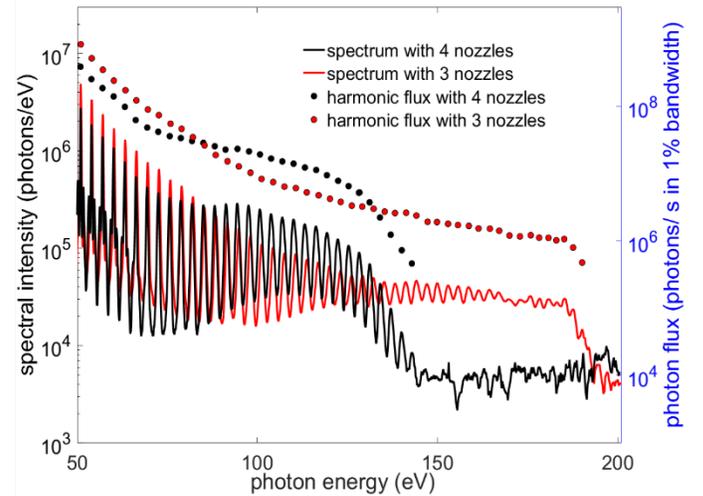

*Figure 3 HHG spectra and photon fluxes (blue axis) generated in helium by the device with 4 evenly distributed nozzles (black) and 3 aperiodic nozzles (red) using a backing pressure of 1 bar.*

respectively. Due to the short propagation distance, we neglected the radiative losses through the hollow waveguide boundaries. The ionization fraction η was estimated by the Yudin *et al.* model [62], leading to $\eta_p$=5.9% at the pulse peak in helium. In the model, we assumed the fundamental pulse to propagate in a medium with a constant ions fraction $\eta_P$ within

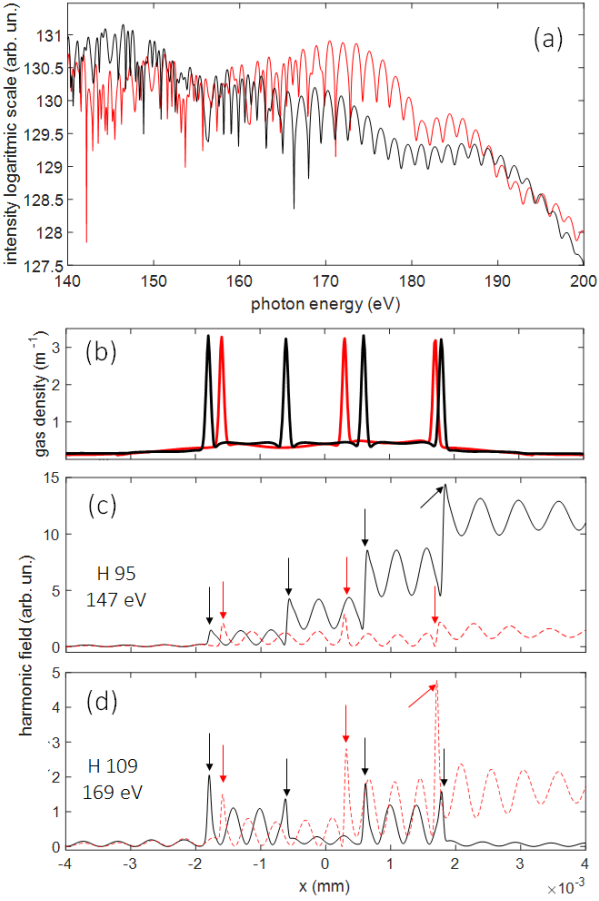

*Figure 5 Computed harmonic field in a modulated gas density. (a) numerical HHG spectra of the three-nozzle (red) and four-nozzle (black) configurations.( b) Numerical gas density profile for the three (red) and four (black) gas jets along the waveguide axis. The axial profile was extrapolated from the 3D numerical calculations. (c), (d) show the growth of the harmonic field along the waveguide axis with the two different gas density distributions for harmonics 95 and 109, respectively. The red and black arrows indicate the position of the gas density peaks.*

the full waveguide length. The phase mismatch for the q-th harmonic is written as [63]:

$$\Delta k_q(x) = -\frac{qu_{11}^2\lambda_0}{4\pi a^2} + k_0 q\left[\left(n_o(\rho) - n_q(\rho)\right)(1-\eta_p) - \frac{\rho}{\rho_{atm}}\eta_p \frac{\omega_p^2}{\omega_0^2}\left(1 - \frac{1}{q^2}\right)\right]$$

Where $\rho_{atm}$, $a$, $u_{11}$, $k_0$, $\omega_0$, and $\omega_p$ represent the gas density (in standard conditions), the waveguide radius, the first zero of the Bessel function $J_0$, the fundamental wavevector, the fundamental frequency, and the plasma frequencies for a fully ionized gas at 1 atm in standard conditons. $n_o(\rho)$ and $n_q(\rho)$ are the refractive index calculated as a function of the local gas density $\rho(x)$.

The harmonic field is computed by solving numerically the growth equation [34]

$$\frac{dE_q}{dx} = -\alpha_q(x)E_q + b(x)e^{i\int_{-x_0}^{x}\Delta k_q(x')dx'}$$

with $\alpha_q$ accounting for the medium absorption as a function of the gas density distribution and $b(x) = iq\omega_0^2\rho(x)d_q(x)(1-\eta_p)/2\varepsilon_0 c_0$. The second term is the polarization field of the q-th order. To isolate the role of the phase-mismatch in the development of the harmonic field, the nonlinear dipole moment $d_q(x)$ is assumed to be unitary and constant for each harmonic order.

Figure 5(a) shows the numerical HHG spectra calculated from the single-atom nonlinear dipole [64] by applying to each harmonic component an amplitude modulation according to the above-described phase-matching model for the two different nozzle distributions, with three (red curve) and four (black curve) jets. The spectra are obtained by averaging the gas density and pulse energy distribution inside the waveguide. Namely, to take into account the non-uniform gas density and pulse intensity as a function of the radial coordinate, the 1D model was applied both along the waveguide axis and on a bundle of lines parallel to the waveguide axis and regularly distributed inside the waveguide volume forming a Cartesian grid.

The model reproduced well the enhancement of the cutoff yield in the three-nozzle configuration compared to the four-nozzle case. To show the nonlinear field growth along the propagation direction as a function of the harmonic order, the results of this simple-man model for harmonics 95 (147 eV) and 109 (169 eV) are shown in 5(c) and 5(d), respectively.

The fields have a characteristic step-like behaviour in correspondence to the gas peaks, overlapped with periodic amplitude modulation. The modulation is due to the residual gas background inside the waveguide and the periodicity is related to the phase cumulated by the harmonic as they propagate in the residual gas. The gas density peaks produce both a change in the harmonic amplitude and a phase change associated with a breaking of the modulation periodicity. Depending on the gas jet position, the local increase of the emitters produces an increase or a decrease in the amplitude of the up-converted field as a result of the phase accumulated by the q-th harmonic inside the gas jet. The four-nozzle scheme allows a favourable field growth in the range 140-150 eV, while it provides a less favourable quasi-phase-matching for harmonics between 160 eV and 200 eV. In this spectral region, the three-nozzle configuration, due to the different inter-nozzle distances, is less selective on the phase, such that an undermined growth of harmonic components in between two jets due to an unfavourable jets distance can be recovered by detuning the distance of the third jet. This makes quasi-phase-matching effects visible over a broader range of harmonics, thus resulting in a higher conversion yield at the cutoff. We estimated that in both the devices, the role of absorption in the gas-jet configurations is negligible, leading to an attenuation of less than 5 % of the harmonic field.

## VI. CONCLUSION AND PERSPETIVES

We demonstrated efficient XUV generation and quasi-phase-matching control, inside microfluidic devices fabricated by the FLICE technique. Our microfluidic source provides a conversion efficiency in line with the state-of-the-art table-top

XUV sources based on HHG. Compared to HHG in static gas cells and gas-filled waveguides, the microfluidic approach offers the unique advantage of enabling the micro-manipulation of the gas distribution within the generation volume, thus allowing accurate control and shaping of the harmonic generation conditions. In this framework, we showed XUV generation by gas density modulation in a periodic and aperiodic micro-jets arrangement. The difference in both the HHG yield and cut-off measured within the two different arrangements demonstrates the possibility to tune phase-matching in a selected portion of the spectrum by the nozzle position. This approach is scalable to multi-bar backing pressures and more complex gas jets architectures, including a higher number of nozzles and customized spatial positioning.

Following the route traced by microfluidics, we foresee the realization of innovative platforms for XUV/X-ray science inspired by the concept of lab-on-a-chip (LOC). Up to date, LOCs are widely exploited for biomedical, chemical, and environmental purposes [65-67], herein mainly applied to the manipulation of liquids. However, they are perfectly suitable for the transport of gases as well. Moreover, despite recent progress in X-ray transport and focusing produced by the advances in diffractive, refractive, guided optics [68, 69], the optical technology behind HHG beamlines remains extremely complex and, thus, restrained to few advanced laboratories.

In this sense, our approach provides a first demonstration of the potential for extending the concept of LOC to X-ray beamlines. By reversing the paradigm of large-scale facilities for X-ray science, we envisage the possibility to realize miniaturized microfluidic experimental stations, taking advantage of the impressive capabilities of FLICE technique. Indeed, such an integrated microfluidic scheme can be extended to encompass more complex optical functionalities, including the interaction with either gas or liquid samples as well as the handling of multiple laser beams by pulse splitting, delay lines, and interferometers [70], thus forwarding the perspective of a new generation of lab-on-a-chip for attosecond spectroscopy and X-ray-based technologies.

## VII. Acknowledgments

We thank Amplitude and Dr. Philippe Demengeot for their support. This project has received funding from the European Unions Horizon 2020 research and innovation program under grant agreement No 964588 (XPIC), by the European Research Council Proof of Concept Grant FESTA (Grant No. 813103) and MSCA-ITN SMART-X (Grant No. 860553), by the Italian Ministry of Research and Education with the projects ELIESFRI Roadmap and PRIN aSTAR (2017RKWTMY), by the Consiglio Nazionale delle Ricerche with the Joint Laboratory ATTOBIO.

## VIII. Author contributions

S.S., R.O. and C.V. conceived the experiments. A.G.C., R.M.V. and G.C. designed and performed the experiments with assistance from M.D., D.F. and M.N. R.M.V. and P.B. fabricated the glass microdevices. M.D., F.F. and L.P. developed the HHG acquisition systems. R.M.V. and A.G.C. and F.B. performed the microuidic simulations. A.F. supervised the gas density simulations. D.F. and A.G.C. performed the numerical HHG calculations. V.T. helped with the design of the devices and the interpretation of the HHG spectra. The first draft of the manuscript and Supplementary Information were written by A.G.C. and R.M.V. All authors discussed the results and contributed to the writing and editing of the manuscript.

## IX. References

[1] F. Calegari, D. Ayuso, A. Trabattoni, L. Belshaw, S. De Camillis, S. Anumula, F. Frassetto, L. Poletto, A. Palacios, P. Decleva, J. B. Greenwood, F. Martín, M. Nisoli, «Ultrafast electron dynamics in phenylalanine initiated by attosecond pulses.» *Science,* vol. 346, n. 6207, pp. 336-339, 2014.

[2] M. Schultze, K. Ramasesha, C. D. Pemmaraju, S. A. Sato, D. Whitmore, A. Gandman, J. S. Prell, L. J. Borja, D. Prendergast, K. Yabana, D. M. Neumark, S. R. Leone, «Attosecond band-gap dynamics in silicon.» vol. 346, n. 6215, pp. 1348-1352, 2014.

[3] T. T. Luu, M. Garg, S. Y. Kruchinin, A. Moulet, M. T. Hassan, E. Goulielmakis, «Extreme ultraviolet high-harmonic spectroscopy of solids.» *Nature,* vol. 521, n. 7553, pp. 498-502, 2015.

[4] A. Sommer, E. M. Bothschafter, S. A. Sato, C. Jakubeit, T. Latka, O. Razskazovskaya, H. Fattahi, M. Jobst, W. Schweinberger, V. Shirvanyan, V. S. Yakovlev, R. Kienberger, Y. K., N. Karpowicz, S. M., F. Krausz, «Attosecond nonlinear polarization and light–matter energy transfer in solids.» *Nature,* vol. 534, n. 7605, pp. 86-90, 2016.

[5] M. T. Hassan, T. T. Luu, A. Moulet, O. Raskazovskaya, P. Zhokhov, M. Garg, N. Karpowicz, A. M. Zheltikov, V. Pervak, F. Krausz, E. Goulielmakis, «Optical attosecond pulses and tracking the nonlinear response of bound electrons.» *Nature,* vol. 530, n. 7588, pp. 66-70, 2016.

[6] D. M. Villeneuve, P. Hockett, M. J. J. Vrakking, H. Niikura, «Coherent imaging of an attosecond electron wave packet.» *Science,* vol. 356, n. 6343, pp. 1150-1153, 2017.

[7] Y. Kobayashi, K. F. Chang, T. Zeng, D. M. Neumark, S. R. Leone, «Direct mapping of curve-crossing dynamics in IBr by attosecond transient absorption spectroscopy.» *Science,* vol. 365, n. 6448, pp. 79-83, 2019.

[8] U. Elu, M. Baudisch, H. Pires, F. Tani, M. H. Frosz, F. Köttig, A. Ermolov, P. S. Russell, J. Biegert, «High average power and single-cycle pulses from a mid-IR optical parametric chirped pulse amplifier.» *Optica,* vol. 4, n. 9, pp. 1024-1029, 2017.

[9] U. Elu, T. Steinle, D. Sànchez, L. Maidment, K. Zawilski, P. Schunemann, U. D. Zeitner, C. Simon-Boisson , J. Biegert, «Table-top high-energy 7 μm OPCPA and 260 mJ Ho: YLF pump laser.» *Optics letters,* vol. 44, n. 13, pp. 3194-3197, 2019.

[10] M. C. Chen, P. Arpin, T. Popmintchev, M. Gerrity, B. Zhang, M. Seaberg, D. Popmintchev, M. M. Murnane, H. C. Kapteyn, «Bright, coherent, ultrafast soft x-ray harmonics spanning the water window from a tabletop light source.» *Physical Review Letters,* vol. 105, n. 17, p. 173901, 2010.

[11] J. Peatross, J. R. Miller, K. R. Smith, S. E. Rhynard, B. W. Pratt, «Phase matching of high-order harmonic generation in helium-and neon-filled gas cells.» *Journal of Modern Optics,* vol. 51, n. 16-18, pp. 2675-2683, 2004.

[12] A. G. Ciriolo, A. Pusala, M. Negro, M. Devetta, F. D., G. Mariani, C. Vozzi, S. Stagira, «Generation of ultrashort pulses by four wave mixing in a gas-filled hollow core fiber.» *Journal of Optics,* vol. 20, n. 12, p. 125503, 2018.

[13] M. Nisoli, S. De Silvestri, O. Svelto, «Generation of high energy 10 fs pulses by a new pulse compression technique.» *Applied Physics Letters,* vol. 68, n. 20, pp. 2793-2795, 1996.

[14] T. Popmintchev, M. C. Chen, D. Popmintchev, P. Arpin, S. Brown, S. Ališauskas, G. Andriukaitis, T. Balčiunas, O. D. Mücke, A. Pugzlys, «Bright coherent ultrahigh harmonics in the keV x-ray regime from mid-infrared femtosecond lasers.» *Science,* vol. 336, n. 6086, pp. 1287-1291, 2012.

[15] C. G. Durfee III, A. R. Rundquist, B. S. C. Herne, M. M. Murnane, H. C. Kapteyn, «Phase matching of high-order harmonics in hollow waveguides.» *Physical Review Letters,* vol. 83, n. 11, p. 2187, 1999.

[16] X. Zhang, A. Lytle, T. Popmintchev, A. Paul, N. Wagner, M. Murnane, H. Kapteyn, I. P. Christov, «Phase matching, quasi-phase matching, and pulse compression in a single waveguide for enhanced high-harmonic generation.» *Optics letters,* vol. 30, n. 15, pp. 1971-1973, 2005.

[17] C. Ding, W. Xiong, T. Fan, D. D. Hickstein, T. Popmintchev, X. Zhang, M. Walls, M. M. Murnane, H. C. Kapteyn, « High flux coherent super-continuum soft X-ray source driven by a single-stage, 10 mJ, Ti: sapphire amplifier-pumped OPA.» *Optics express,* vol. 22, n. 5, pp. 6194-6202, 2014.

[18] E. Constant, D. Garzella, P. M. E. Breger, C. Dorrer, C. Le Blanc, F. Salin, P. Agostini, «Optimizing high harmonic generation in absorbing gases: Model and experiment.» *Physical Review Letters,* vol. 82, n. 8, p. 1668, 1999.

[19] G. Porat, C. M. Heyl, S. B. Schoun, C. Benko, D. N., K. L. Corwin, J. Ye, «Phase-matched extreme-ultraviolet frequency-comb generation.» *Nature Photonics,* vol. 12, n. 7, pp. 387-391, 2018.

[20] E. A. Gibson, A. Paul, N. Wagner, D. Gaudiosi, S. Backus, I. P. Christov, A. Aquila, E. M. Gullikson, D. T. Attwood, M. M. Murnane, H. C. Kapteyn, «Coherent soft x-ray generation in the water window with quasi-phase matching.» *Science,* vol. 302, n. 5642, pp. 95-98, 2003.

[21] A. Paul, R. A. Bartels, R. Tobey, H. Green, S. Weiman, I. P. Christov, M. M. Murnane, H. C. Kapteyn, S. Backus, «Quasi-phase-matched generation of coherent extreme-ultraviolet light.» *Nature,* vol. 421, n. 6918, pp. 51-54, 2003.

[22] B. Dromey, M. Zepf, M. Landreman, S. M. Hooker, «Quasi-phasematching of harmonic generation via multimode beating in waveguides.» *Optics express,* vol. 15, n. 13, pp. 7894-7900, 2007.

[23] I. P. Christov, «Propagation of ultrashort pulses in gaseous medium: breakdown of the quasistatic approximation.» *Optics express,* vol. 6, n. 2, pp. 34-39, 2000.

[24] I. P. Christov, «Control of high harmonic and attosecond pulse generation in aperiodic modulated waveguides.» *JOSA B,* vol. 18, n. 12, pp. 1877-1881, 2001.

[25] A. Bahabad, O. Cohen, M. M. Murnane, H. C. Kapteyn, «Quasi-periodic and random quasi-phase matching of high harmonic generation.» *Optics letters,* vol. 33, n. 17, pp. 1936-1938, 2001.

[26] V. Tosa, V. S. Yakovlev e F. Krausz, «Generation of tunable isolated attosecond pulses in multi-jet systems.» *New Journal of Physics,* vol. 10, n. 2, p. 025016, 2008.

[27] K. C. Vishnubhatla, N. Bellini, R. Ramponi, G. Cerullo, R. Osellame, «Shape control of microchannels fabricated in fused silica by femtosecond laser irradiation and chemical etching.» *Optics express,* vol. 17, n. 10, pp. 8685-8695, 2009.

[28] R. Osellame, H. J. Hoekstra, G. Cerullo, M. Pollnau, «Femtosecond laser microstructuring: an enabling tool for optofluidic lab-on-chips.» *Laser and Photonics Reviews,* vol. 5, n. 3, pp. 442-463, 2011.

[29] F. He, J. Lin, Y. Cheng, «Fabrication of hollow optical waveguides in fused silica by three-dimensional femtosecond laser micromachining.» *Applied Physics B,* vol. 105, n. 2, pp. 379-384, 2011.

[30] R. Martínez Vázquez, A. G. Ciriolo, G. Crippa, V. Tosa, F. Sala, M. Devetta, C. Vozzi, S. Stagira, R. Osellame, «Femtosecond laser micromachining of integrated glass devices for high-order harmonic generation.» *International Journal of Applied Glass Science,* vol. 13, n. 2, pp. 162-170, 2022.

[31] A. G. Ciriolo, R. Martínez Vázquez, V. Tosa, A. Frezzotti, G. Crippa, M. Devetta, D. Faccialà, F. Frassetto, L. Poletto, A. Pusala, C. Vozzi, R. Osellame, S. Stagira, «High-order harmonic generation in a microfluidic glass device.» *Journal of Physics: Photonics,* vol. 2, n. 2, p. 024005, 2020.

[32] T. Auguste, B. Carré, P. Salières, «Quasi-phase-matching of high-order harmonics using a modulated atomic density.» *Physical Review A,* vol. 76, n. 1, p. 011802, 2007.

[33] J. Seres, V. S. Yakovlev, E. Seres, C. Streli, P. Wobrauschek, C. Spielmann, F. Krausz, «Coherent superposition of laser-driven soft-X-ray harmonics from successive sources.» *Nature Physics,* vol. 3, n. 12, pp. 878-883, 2007.

[34] I. Hadas, A. and Bahabad, «Periodic density modulation for quasi-phase-matching of optical frequency conversion is inefficient under shallow focusing and constant ambient pressure.» *Optics letters,* vol. 41, n. 17, pp. 4000-4003, 2016.

[35] L. D. Landau, E. M. Lifshitz, Fluid Mechanics, Pergamon Press, 1987.

[36] C. Cercignani, Rarefied Gas Dynamics, Cambridge University Press, 2000.

[37] COMSOL Multiphysics, *v. 5.4. www.comsol.com. COMSOL AB,* Stockholm: Sweden.

[38] E. A. J. Marcatili, R. A. Schmeltzer, «Hollow metallic and dielectric waveguides for long distance optical transmission and lasers.» *The Bell System Technical Journal,* vol. 43, n. 4, pp. 1783-1809, 1964.

[39] L. Poletto, G. Tondello, P. Villoresi, «High-order laser harmonics detection in the EUV and soft x-ray spectral regions.» *Review of Scientific Instruments,* vol. 72, n. 7, pp. 2868-2874, 2001.

[40] The Center for X-Ray Optics (CXRO), https://henke.lbl.gov/optical_constants/.

[41] L. Poletto, G. Naletto, G. Tondello, «Grazing-incidence flat-field spectrometer for high-order harmonic diagnostics.» *Optical Engineering,* vol. 40, n. 2, pp. 178-185, 2001.

[42] [Online]. Available: https://www.princetoninstruments.com/products/pixis-family/pixis-xo.

[43] C. Ding, W. Xiong, T. Fan, D. D. Hickstein, T. Popmintchev, X. Zhang, M. Walls, M. Murnane, H. C. Kapteyn, «High flux coherent super-continuum soft X-ray source driven by a single-stage, 10mJ, Ti: sapphire amplifier-pumped OPA.» *Optics express,* vol. 22, n. 5, pp. 6194-6202, 2014.

[44] S. Hädrich, A. Klenke, J. Rothhardt, M. Krebs, A. Hoffmann, O. Pronin, V. Pervak, J. Limpert, A. Tünnermann, « High photon flux table-top coherent extreme-ultraviolet source.» *Nature Photonics,* vol. 8, n. 10, pp. 779-783, 2014.

[45] S. Hädrich, M. Krebs, A. Hoffmann, A. Klenke, J. Rothhardt, J. Limpert, A. Tünnermann, «Exploring new avenues in high repetition rate table-top coherent extreme ultraviolet sources.» *Light: Science & Applications,* vol. 4, n. 8, pp. e320-e320, 2015.

[46] J. Rothhardt, M. Krebs, S. Hädrich, S. Demmler, J. Limpert, A. Tünnermann, «Absorption-limited and phase-matched high harmonic generation in the tight focusing regime.» *New Journal of Physics,* vol. 16, n. 3, p. 033022, 2014.

[47] E. J. Takahashi, Y. Nabekawa, K. Midorikawa, «Low-divergence coherent soft x-ray source at 13 nm by high-order harmonics.» *Applied physics letters,* vol. 84, n. 1, pp. 4-6, 2004.

[48] E. J. Takahashi, Y. Nabekawa, H. Mashiko, H. Hasegawa, A. Suda, K. Midorikawa, «Generation of strong optical field in soft X-ray region by using high-order harmonics.» *IEEE journal of selected topics in quantum electronics,* vol. 10, n. 6, pp. 1315-1328, 2004.

[49] P. Rudawski, C. M. Heyl, F. Brizuela, J. Schwenke, A. Persson, E. Mansten, R. Rakowski, L. Rading, F. Campi, B. Kim, P. Johnsson, A. L'Huillier, «A high-flux high-order harmonic source.» *Review of Scientific Instruments,* vol. 84, n. 7, p. 073103, 2013.

[50] A. Cingöz, D. C. Yost, T. K. Allison, A. Ruehl, M. E. Fermann, I. Hartl, J. Ye, «Direct frequency comb spectroscopy in the extreme ultraviolet.» *Nature,* vol. 482, n. 7383, pp. 68-71, 2012.

[51] G. Porat, C. M. Heyl, S. B. Schoun, C. Benko, N. Dörre, K. L. Corwin e J. Ye, «Phase-matched extreme-ultraviolet frequency-comb generation.» *Nature Photonics,* vol. 12, n. 7, pp. 387-391, 2018.

[52] I. Pupeza, S. Holzberger, T. Eidam, H. Carstens, D. Esser, J. Weitenberg, P. Rußbüldt, J. Rauschenberger, J. Limpert, T. Udem, A. Tünnermann, T. W. Hänsch, A. Apolonski, F. Krausz, E. Fill, «Compact high-repetition-rate source of coherent 100 eV radiation.» *Nature Photonics,* vol. 7, n. 8, pp. 608-612, 2013.

[53] T. Saule, S. Heinrich, J. Schötz, N. Lilienfein, M. Högner, O. deVries, M. Plötner, J. Weitenberg, D. Esser, J. Schulte, P. Russbueldt, J. Limpert, M. Kling, U. Kleineberg, I. Pupeza, « High-flux ultrafast extreme-ultraviolet photoemission spectroscopy at 18.4 MHz pulse repetition rate.» *Nature,* vol. 10, n. 1, pp. 1-10, 2019.

[54] L. Barreau, A. D. Ross, S. Garg, P. M. Kraus, D. M. Neumark, S. R. Leone, «Efficient table-top dual-wavelength beamline for ultrafast transient absorption spectroscopy in the soft X-ray region.» *Scientific reports,* vol. 10, n. 1, pp. 1-9, 2020.

[55] E. J. Takahashi, T. Kanai, K. L. Ishikawa, Y. Nabekawa, K. Midorikawa, «Coherent water window x ray by phase-matched high-order harmonic generation in neutral media.» *Physical Review Letters,* vol. 101, n. 25, p. 253901, 2008.

[56] M. Gebhardt, T. Heuermann, R. Klas, C. Liu, A. Kirsche, M. Lenski, Z. Wang, C. Gaida, J. Antonio-Lopez, A. Schülzgen, R. Amezcua-Correa, J. Rothhardt, J. Limpert, «Bright, high-repetition-rate water window soft X-ray source enabled by nonlinear pulse self-compression in an antiresonant hollow-core fibre.» *Light: Science & Applications,* vol. 10, n. 1, pp. 1-7, 2021.

[57] M. C. Chen, P. Arpin, T. Popmintchev, M. Gerrity, B. Zhang, M. Seaberg, D. Popmintchev, M. Murnane, H. C. Kapteyn, «Bright, coherent, ultrafast soft x-ray harmonics spanning the water window from a tabletop light source.» *Physical review letters,* vol. 105, n. 17, p. 173901, 2010.

[58] G. Fan, K. Legare, V. Cardin, X. Xie, E. Kaksis, G. Andriukaitis, A. Pugzlys, B. Schmidt, J. Wolf, M. Hehn, G. Malinowski, B. Vodungbo, E. Jal, J. Luning, N. Jaouen, Z. Tao, A. Baltuska, F. Legare, T. Balciunas, «Time-resolving magnetic scattering on rare-earth

ferrimagnets with a bright soft-X-ray high-harmonic source.» *arXiv preprint,* p. 1910.14263, 2019.

[59] Y. Fu, K. Nishimura, R. Shao, A. Suda, K. Midorikawa, P. Lan, E. J. Takahashi, «High efficiency ultrafast water-window harmonic generation for single-shot soft X-ray spectroscopy.» *Communications Physics,* vol. 3, n. 1, pp. 1-10, 2020.

[60] A. S. Johnson, D. R. Austin, D. A. Wood, C. Brahms, A. Gregory, K. B. Holzner, S. Jarosch, E. W. Larsen, S. S. Parker, C. S. Strüber, P. Ye, J. W. G. Tisch, J. P. Marangos, «High-flux soft x-ray harmonic generation from ionization-shaped few-cycle laser pulses.» *Science advances,* vol. 4, n. 5, p. eaar3761, 2018.

[61] J. Rothhardt, S. Hädrich, A. Klenke, S. Demmler, A. Hoffmann, T. Gotschall, T. Eidam, M. Krebs, J. Limpert, A. Tünnermann, «53 W average power few-cycle fiber laser system generating soft x rays up to the water window.» *Optics letters,* vol. 39, n. 17, pp. 5224-5227, 2014.

[62] G. L. Yudin, M. Y. Ivanov, «Nonadiabatic tunnel ionization: Looking inside a laser cycle.» *Physical Review A,* vol. 64, n. 1, p. 013409, 2001.

[63] C. G. Durfee III, A. R. Rundquist, S. Backus, C. Herne, M. M. Murnane, H. C. Kapteyn, «Phase matching of high-order harmonics in hollow waveguides.» *Physical Review Letters,* vol. 83, n. 11, p. 2187, 1999.

[64] M. Lewenstein, P. Balcou, M. Y. Ivanov, A. L'Huillier, P. B. Corkum, «Theory of high-harmonic generation by low-frequency laser fields.» *Physical Review A,* vol. 49, n. 3, p. 2117, 1994.

[65] R. Pol, F. Céspedes, D. Gabriel, M. Baeza, «Microfluidic lab-on-a-chip platforms for environmental monitoring.» *Trends in Analytical Chemistry,* vol. 95, pp. 62-68, 2017.

[66] H. Shi, K. Nie, B. Dong, M. Long, H. Xu, Z. Liu, «Recent progress of microfluidic reactors for biomedical applications.» *Chemical Engineering Journal,* vol. 361, pp. 635-650, 2019.

[67] X. Cheng, M. D. Ooms, D. Sinton, «Biomass-to-biocrude on a chip via hydrothermal liquefaction of algae.» *Lab on a Chip,* vol. 16, n. 2, pp. 256-260, 2016.

[68] T. Salditt, M. Osterhoff, « X-ray Focusing and Optics.» *Nanoscale Photonic Imaging,* pp. 71-124, 2020.

[69] T. Salditt, S. Hoffmann, M. Vassholz, J. Haber, M. Osterhoff, J. Hilhorst, «X-ray optics on a chip: guiding x rays in curved channels.» *Physical review letters,* vol. 105, n. 20, p. 203902, 2015.

[70] A. Crespi, Y. Gu, B. Ngamsom, H. J. Hoekstra, C. Dongre, M. Pollnau, R. Ramponi, H. H. van den Vlekkert, P. Watts, G. Cerullo, R. Osellame, «Three-dimensional Mach-Zehnder interferometer in a microfluidic chip for spatially-resolved label-free detection.» *Lab on a Chip,* vol. 10, n. 9, pp. 1167-1173, 2010.